\documentclass[aps,pra,twocolumn,superscriptaddress,showpacs,float,nofootinbib]{revtex4-1}

\usepackage[squaren]{SIunits}
\usepackage[utf8]{inputenc}
\usepackage[english]{babel}
\usepackage{graphicx}
\usepackage{amsmath,amsfonts}
\usepackage{color}
\usepackage[normalem]{ulem}

\begin{document}

\title{Surface-modified Wannier-Stark states in a 1D optical lattice}

\author{A. Maury}
\affiliation{Laboratoire Kastler Brossel, UPMC-Sorbonne Universités, CNRS, ENS-PSL-Research University, Collège de France; 4 place Jussieu, F-75005 Paris, France}
\email{maury@lkb.upmc.fr}
\author{M. Donaire}
\affiliation{Laboratoire Kastler Brossel, UPMC-Sorbonne Universités, CNRS, ENS-PSL-Research University, Collège de France; 4 place Jussieu, F-75005 Paris, France}
\author{M.-P. Gorza}
\affiliation{Laboratoire Kastler Brossel, UPMC-Sorbonne Universités, CNRS, ENS-PSL-Research University, Collège de France; 4 place Jussieu, F-75005 Paris, France}
\author{A. Lambrecht}
\affiliation{Laboratoire Kastler Brossel, UPMC-Sorbonne Universités, CNRS, ENS-PSL-Research University, Collège de France; 4 place Jussieu, F-75005 Paris, France}
\author{R. Guérout}
\affiliation{Laboratoire Kastler Brossel, UPMC-Sorbonne Universités, CNRS, ENS-PSL-Research University, Collège de France; 4 place Jussieu, F-75005 Paris, France}

\begin{abstract}
We study the energy spectrum of atoms trapped in a vertical 1D optical lattice in close proximity to a reflective surface. We
propose an effective model to describe the interaction between the atoms and the surface at any distance. Our model includes the
long-range Casimir-Polder potential together with a short-range Lennard-Jones potential, which are considered non-perturbatively
with respect to the optical lattice potential. We find an intricate energy spectrum which contains a pair of loosely-bound states
localized close to the surface in addition to a surface-modified Wannier-Stark ladder at long distances. Atomic interferometry
involving those loosely-bound atom-surface states is proposed to probe the adsorption dynamics of atoms on mirrors.
\end{abstract}

\pacs{}

\maketitle

\section{Introduction}

Trapping and manipulating cold neutral atoms in an optical lattice offers high control over the atomic locations and robust
quantum coherence on the dynamics of the atomic states. These properties make of an optical lattice an ideal system for
applications in metrology \cite{Clade2006,Takamoto2005} and in precision measurements of the interactions between the atoms and the
environment \cite{Impens2014}. It is to the latter that the FORCA-G project applies \cite{Wolf2007}. In particular, the
FORCA-G experiment aims at performing high precision measurements of the electromagnetic and gravitational interactions between a
neutral atom and a massive dielectric surface. Ultimately, it aims at establishing accurate constraints in the search of
hypothetical deviations from the Newtonian law of gravitation at short length scales, reason why an accurate knowledge of the
electromagnetic interaction is also needed. It is on the electromagnetic interaction that we concentrate in this article.

\begin{figure}[h!]
\includegraphics[width=0.4\textwidth]{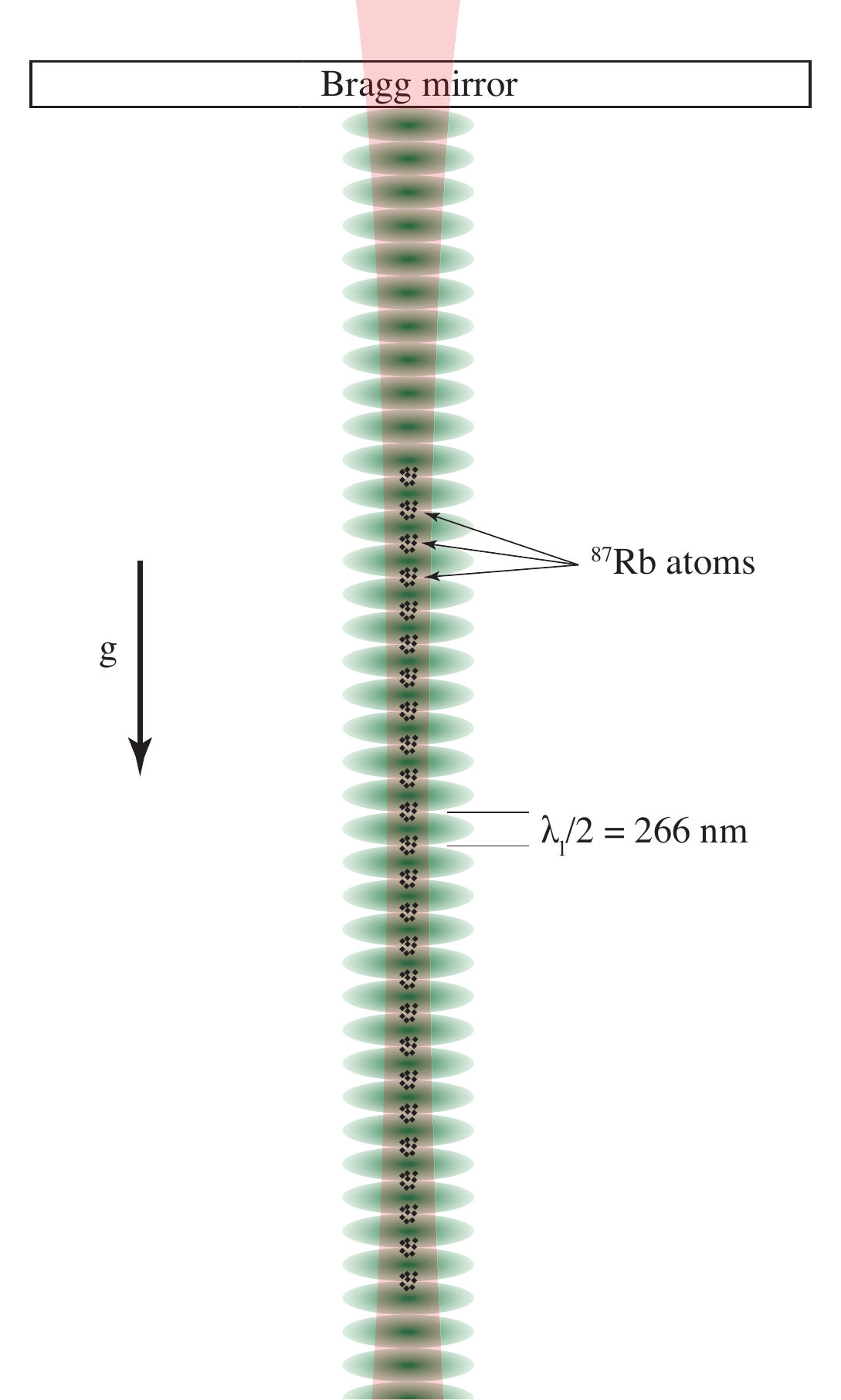}
\caption{Scheme of the experimental device. Cold $^{87}$Rb atoms are trapped in a blue-detuned vertical optical lattice. An infrared laser at $\lambda=1064$ nm assures the transverse confinement. A pair of contrapropagating Raman lasers at $\lambda=780$ nm (not shown) drives the transitions between lattice sites.}
\label{fig:expScheme}
\end{figure}
\begin{figure}[h!]
\includegraphics[width=0.44\textwidth]{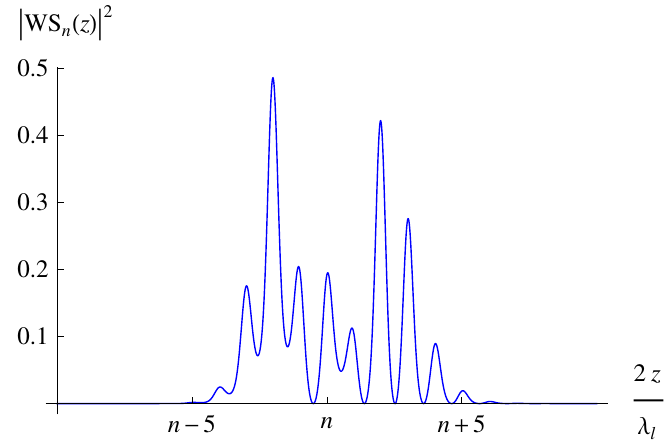}\\
\includegraphics[width=0.44\textwidth]{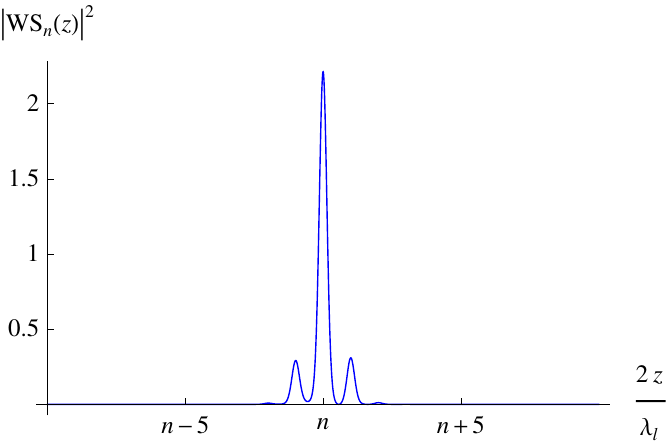}
\caption{Profile of the squared-norm of the wavefunction of the $n^{\text{th}}$ WS state for different values of the optical 
depth, $U=3$ $E_r$ (upper figure) and $U=9$ $E_r$ (lower figure).}
\label{fig:regWS}
\end{figure}

In the setup of FORCA-G atoms of $^{87}$Rb are trapped in a vertical optical lattice by the potential generated by the standing
waves of a laser source of wavelength $\lambda_{l}=532$ nm, which reflect off a Bragg mirror
[see Fig.~\ref{fig:expScheme}]. The optical potential takes the periodic form 
\begin{equation}
	\label{eq:optPot}
	V_{\text{op}}(z)=U(1-\cos{2k_{l}z)}/2
\end{equation}
where
$k_{l}=2\pi/\lambda_{l}$, $z$ is the vertical distance relative to the surface position and $U$ is the optical depth which
depends on the laser intensity. In addition, the uniform Earth gravitational field creates a linear potential
\begin{equation}
	\label{eq:gravPot}
	V_{\text{g}}(z)=-mgz
\end{equation}
with $m$ being the atomic mass and $g$ being the gravitational acceleration. Disregarding the atom-mirror
interaction, the spectrum which results from the addition of the optical and gravitational potentials consists of a ladder of
quasi-stable states referred to as Wannier-Stark (WS) states. The WS eigenstates are localized around the equilibrium points
$z_{n}=n\lambda_{l}/2$, $n$ being an integer, and are uniformly distributed along the energy spectrum at constant intervals
$mg\lambda_{l}/2=h\nu_{B}$. In this expression $\nu_{B}$ is referred to as Bloch frequency, and the degree of localization is
determined by the relative optical depth with respect to the recoil energy, $U/(\hbar^{2}k_{l}^{2}/2m)=U/E_r$ [see
Fig.~\ref{fig:regWS}].

In addition to $V_{\text{op}}(z)$ and $V_{\text{g}}(z)$, the neutral atoms interact with the surface through the mutual coupling of
their charge fluctuations to the vacuum fluctuations of the electromagnetic field. This interaction is known generically as
Casimir-Polder (CP) interaction \cite{Casimir1948,Lifshitz1956}. At zero temperature its strength depends generally on the
dielectric properties of the surface, the state of the atom, and the distance between them.

The \emph{modus operandi} of FORCA-G consists of a sequence of pulses generated by Raman lasers and microwaves which are used to
create an atomic interferometer. The pulses drive the atoms through a coherent superposition of low-lying Zeeman sublevels at
different lattice sites~\cite{Wolf2007}. The CP interaction induces a phase shift on the atoms which depends strongly on the
distance of the atoms to the surface and slightly on the internal atomic states. The phase shift accumulated by the atomic wave
function throughout the sequence of pulses is finally measured by atomic interferometry techniques. If the atoms are made to
oscillate between lattice sites far from the surface, the CP-induced shift is additive. Therefore, once substracted the phase shift
associated to the passage through different WS levels, which is characteristic of the interferometer scheme, the remaining phase is
the CP-induced shift we are interested in.

The latter applies to the case where the CP interaction is small with respect to the optical potential depth, so that it can be
treated as a perturbation to the potential $V_{\text{op}}(z)+V_{\text{g}}(z)$ and hence to the original WS states. This takes place
at separation distances of the order of microns, at which the perturbative development of Refs.~\cite{Messina2011,Pelisson2013}
applies. On the contrary, at submicrometer distances and beyond the perturbative regime, it was already noticed in
Ref.~\cite{Messina2011} that the CP corrections to the original WS energies diverge. This is specially relevant to the purposes of
the FORCA-G project, as deviation from Newtonian gravity are expected to occur at submicrometer distances. Therefore, a precise
knowledge of the CP interaction at this length scale as well as an accurate description of the spatial distribution of the atomic
wave function are crucial in order to detect those gravitational effects. In reference~\cite{Messina2011} the authors apply a
regularization scheme for the CP potential based on the assumption that the minimum distance of the atom to the surface is
determined by the atomic radius. However, it is found there that the resultant corrections strongly depend on this radius. Thus,
non-reliable results were obtained.

It is the main purpose of the present article to develop a non-perturbative approach to this problem in order to determine
accurately the energy spectrum and the profile of the atomic states at submicrometer distances. To this end, we model the
short-range interaction between the atom and the surface by a Lennard-Jones potential which features the adsorption of the atoms on
the surface. We find that, in addition to slightly modified WS states, the resultant spectrum contains a number of loosely-bound
atom-surface states whose properties depend critically on the parameters of the Lennard-Jones potential. Nonetheless, independent
measurements can be performed to determine the unknowns of such potential.


The remainder of the article is organized as follows. In section~\ref{sec:atom_surface_potential}, we present the features of the
potential modelling the interaction between the atom and the surface. In section~\ref{sec:surface_modified_wannier_stark_states},
we show that the overall effect of the surface leads to a complex energy spectrum significantly departing from the usual
Wannier-Stark states. We conclude by calculating a typical atomic interferometry spectrum obtained using stimulated Raman
transitions between those surface-modified Wannier-Stark states.

\section{The atom-surface potential}
\label{sec:atom_surface_potential}

In addition to the optical potential described in the precedent section, the atoms interact with the mirror through the
electromagnetic field. Quite generally, this interaction is made of two distinct components, namely a short-range and a long-range
potentials. The short-range potential results from the spatial overlap between the electronic clouds of the atoms and the surface at
sub-nanometer distances. In turn, this potential depends on the precise profile of the electronic density distribution, which is
difficult to determine both experimentally and theoretically. Hence, a parametrization scheme is required for the short-range
potential. In contrast, the long-range potential originates from the mutual coupling of the charges within the atoms and the
currents in the mirror to the fluctuating electromagnetic field. This is the so-called Casimir-Polder potential, which is computed
in the electric dipole approximation at second order in stationary perturbation theory.

In the framework of the scattering theory~\cite{Lambrecht2006}, the Casimir-Polder potential between a flat
mirror in the $(xy)$ plane and an atom in the ground state separated by a distance $z$, at temperature $T$, is given
by~\cite{Dufour2013}
\begin{align}
	\label{eq:CPeq}
V_{\text{s}}^{\text{CP}}(z)=k_BT\sum_{n}{}^{'} \frac{\xi_{n}^{2}}{c^{2}}\frac{\alpha(i \xi_{n})}{4\pi\epsilon_0}\int^{\infty}_0 \frac{\text{d}^2\mathbf{k}}{\kappa}e^{-2\kappa z}\nonumber \\
\times\left(\rho^{TE}-\left(1+\frac{2\kappa^2 c^{2}}{\xi_{n}^{2}}\right)\rho^{TM}\right)
\end{align}
with $\mathbf{k}^2=k_x^2+k_y^2$, $\kappa=\sqrt{\mathbf{k}^2+\xi_{n}^{2}/c^{2}}$ and the sum runs over Matsubara frequencies
$\xi_n = 2 \pi n k_B T / \hbar$. In this equation $\rho^{TE}$ and $\rho^{TM}$ are the reflection coefficients of
the mirror for the $TE$ and $TM$ polarizations, respectively, and $\alpha(i\xi)$ is the polarizability of
a $^{87}$Rb atom in its ground state evaluated at imaginary frequencies~\cite{Derevianko2010},
\begin{equation}
\alpha(i\xi)=\frac{2}{\hbar}\sum_j{\frac{\omega_{jg}d_{jg}^2}{\omega_{jg}^2+\xi^2}},
\end{equation} 
where $\omega_{jg}=\omega_j-\omega_g$ and $d_{jg}$ are respectively the transition frequency and the electric dipole
matrix element between the states $\left|j\right\rangle$ and $\left|g\right\rangle$. 

Concerning the optical properties of the mirror used in the FORCA-G experiment, its design is such that it is nearly transparent at
780 nm and 1064 nm while it is reflective at 532 nm. It is a Bragg mirror formed by alternating layers of SiO$_{2}$ and
Ta$_{2}$O$_{5}$. Its reflection coefficients $\rho^{TE}$ and $\rho^{TM}$ are obtained using standard transfer matrix theory. Let us
define first by $T_{i}$ the transfer matrix associated to the transmition through the interface between the layers $i$ and $i+1$, as
well as to the propagation throughout the layer $i+1$ of width $d_{i+1}$. It relates the field on the left of the layer $i$ to the
field on the right, and reads~\cite{Ingold2015}
\begin{equation}
  \label{eq:transfer}
  T_{i}=\frac{1}{\bar{t}_i}
  \begin{pmatrix}
  	  t_i\bar{t}_i-r_i\bar{r}_i & \bar{r}_i \\
	  -r_i & 1
  \end{pmatrix}\begin{pmatrix}
  	e^{i k_{z} d_{i+1}} & 0 \\
	0 & e^{-i k_{z} d_{i+1}}
  \end{pmatrix}
\end{equation}
In this equation, $r_i$ and $t_i$ are the Fresnel amplitudes from medium $i$ to medium $i+1$. The barred quantities are the reciprocal
amplitudes from medium $i+1$ to medium $i$ and $k_{z}$ is the $z$-component of the wavevector in medium $i+1$. The
transfer matrix of the Bragg mirror, $\mathbb{T}$, is the product of the transfer matrices of all the layers
$\mathbb{T}=\prod_{i}T_{i}$ and reads
\begin{equation}
  \label{eq:totalTransfer}
  \mathbb{T}=\frac{1}{\bar{\tau}}
  \begin{pmatrix}
  	  \tau\bar{\tau}-\rho\bar{\rho} & \bar{\rho} \\
	  -\rho & 1
  \end{pmatrix}
\end{equation}
from which the total reflection amplitude reads $\rho=-[\mathbb{T}]_{21}/[\mathbb{T}]_{22}$.

\begin{figure}[h]
  \centering
    \includegraphics[width=0.45\textwidth]{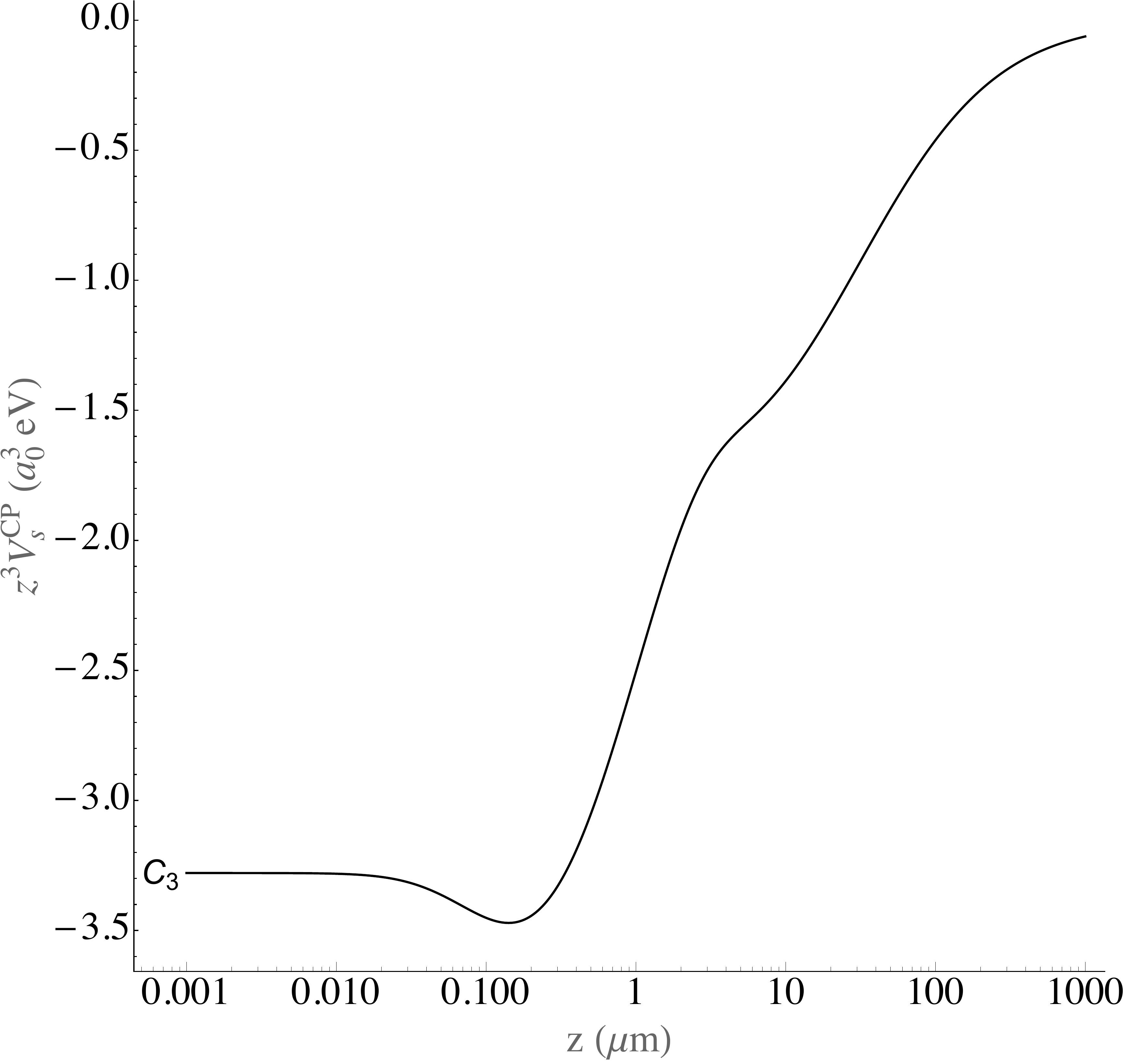}
  \caption{The Casimir-Polder potential calculated between a Rubidium atom and a SiO$_{2}$–Ta$_{2}$O$_{5}$ Bragg mirror as a function of the distance $z$. The value of the van der Waals coefficient $C_{3}$ is indicated.}
  \label{fig:Figures_PotentialCP}
\end{figure}

We show in Figure~\ref{fig:Figures_PotentialCP} the Casimir-Polder potential calculated using Equation~\eqref{eq:CPeq} for a temperature $T=300$ K.
The potential is scaled with $z^{3}$, the third power of the atom-surface distance, in order to emphasize the non-retarded
van der Waals regime characterized by its coefficient $C_{3}\approx 3.28$ a$_{0}^{3}$eV.

As for the short-range potential, we parametrize it using a $12-3$ Lennard-Jones form,\footnote{In surface
science, a $9-3$ Lennard-Jones potential is also often used as it arises as pairwise summation of $12-6$ Lennard-Jones atom-atom
interactions.}
\begin{equation}
  \label{eq:Vsurf}
  V_{\text{s}}^{\text{LJ}}(z)=\frac{D}{3}\left(\left(\frac{z_{0}}{z}\right)^{12}-4\left(\frac{z_{0}}{z}\right)^3\right),
\end{equation}
which is characterized by a well depth $D$ and an equilibrium distance $z_{0}$ which correspond to the energy and distance from
the surface of an adsorbed atom, respectively. 
Continuity of the atom-surface potential demands that $V_\text{s}^\text{LJ}(z)$ and $V_\text{s}^\text{CP}(z)$ smoothly
merge at some intermediate distance $z_m$.
This implies that $D$ and $z_{0}$ are no longer independent but are related by the
equation
\begin{equation}
  \label{eq:related}
  \frac43 D z_{0}^{3}=C_{3}
\end{equation}
where $C_{3}$ is the van der Waals coefficient in the Casimir-Polder potential. With this condition between the parameters $D$ and $z_{0}$ in the Lennard-Jones potential, the matching distance $z_m$ is chosen where both potentials $V_{\text{s}}^\text{LJ}(z)$ and $V_{\text{s}}^\text{CP}(z)$ behave in $z^{-3}$ and leads to the total surface potential $V_\text{s}(z)$:
\begin{equation}
	\label{eq:totVs}
	V_\text{s}(z)=V_{\text{s}}^{\text{LJ}}(z)\Theta(z_m-z)+V_{\text{s}}^{\text{CP}}(z)\Theta(z-z_m)
\end{equation}
where $\Theta(z)$ is the Heaviside function.

The form used for
$V_{\text{s}}^{\text{LJ}}(z)$ is merely of a physisorption-type, and hence expected to underestimate the adsorption
energy. For instance, for an equilibrium distance $z_{0}=2.3$ \AA\, we find $D\approx 30$ meV to be compared with a value of
$\approx 350$ meV from a recent density functional theory calculation~\cite{Sedlacek2015}. 
As a matter of fact, the parameters of the short-range potential carry the largest uncertainty in our calculation. An accurate
determination of this part of the potential would require extensive \emph{ab initio} calculations up to distances of the order of
the nanometers which are beyond the scope of this work. Alternatively, the parameters $D$ and $z_{0}$ can be determined experimentally. Be that as it may, we
will study in the next section the influence of our results upon the parameters of the Lennard-Jones model.

\section{Surface-modified Wannier-Stark states} 
\label{sec:surface_modified_wannier_stark_states}

\begin{figure}[h]
  \centering
    \includegraphics[width=.45\textwidth]{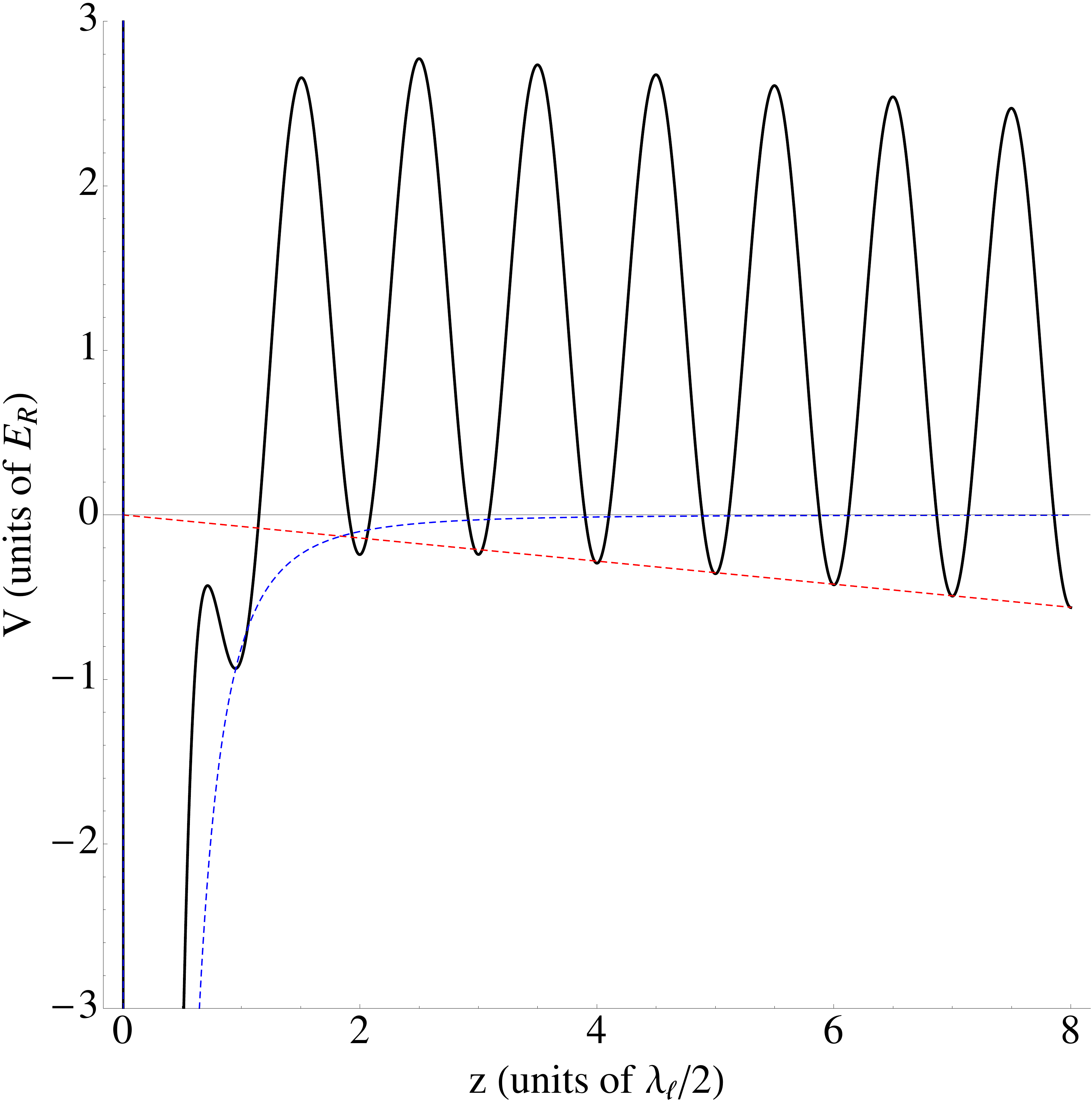}
  \caption{(color online) Potential $V(z)$ in units of the recoil energy $E_{r}$ shown as the black curve. The dashed blue and red curve are, respectively, the surface potential $V_{\text{s}}(z)$ and the gravitational potential $-mgz$.}
  \label{fig:Figures_Potential}
\end{figure}

In the following and unless otherwise stated, we will refer to the distance $z$ to the surface in units of
$\lambda_{l}/2=266$ nm and the energies in units of the recoil energy $E_{r}=\frac{\hbar^{2} k_{l}^{2}}{2 m}\approx 5.37
\times 10^{-30}$ J for a Rubidium atom. The surface-modified Wannier-Stark states (SMWSS) are solutions
of the time-independent Schrödinger equation
\begin{align}
  \label{eq:1DSchrod}
  -\frac{\hbar^{2}}{2m}\frac{\text{d}^{2}\psi_{n}(z)}{\text{d}z^{2}}+V(z)\psi_{n}(z)=E_{n}\psi_{n}(z),\\
  \text{with }V(z)=V_{\text{s}}(z)+V_{\text{g}}(z)+V_{\text{op}}(z).
\end{align}
In the situation where the mirror is \emph{above} the atoms, the potential $V(z)$ is not bounded from below so that all states are
rigorously Siegert states~\cite{Siegert1939}. That corresponds to the situation where the atoms could ultimately "fall from the
optical lattice". Nevertheless, it has been shown in reference~\cite{Pelisson2013} that lifetimes are of the order of 10$^{14}$ s for the
first Bloch band and hence they can be considered stable for any experimental realization. We show in Figure~\ref{fig:Figures_Potential} the potential $V(z)$ for an optical depth $U=3 $
$E_{r}$.
At $z\approx 2$, the magnitudes of the gravitational and the Casimir-Polder potentials are similar. As a result, the very
first optical well is strongly influenced by the surface to the point of becoming weakly bounding. Note that the minimum of the
Lennard-Jones part of the surface potential have very different orders of magnitude, both in binding energy ($D\approx
10^{9} E_{r}$) and equilibrium distance ($z_{0}=2.3$ \AA $\,\approx 10^{-3}\lambda_{l}/2)$, reason why it does not appear in
Figure~\ref{fig:Figures_Potential}.

The SMWSS $\psi_{n}(z)$ are conveniently characterized according to their mean distance to the surface $\langle z \rangle$,
\begin{equation}
  \label{eq:meanZ}
  \langle z \rangle =\frac{\langle \psi_{n}|z|\psi_{n}\rangle}{\langle \psi_{n}|\psi_{n}\rangle}.
\end{equation}
We show in Table~\ref{tab:smwss} values of the mean distance $\langle z \rangle$ and the energy intervals for the first few SMWSS calculated for an optical depth $U=3$ $E_{r}$ and $z_{0}=2.3$ \AA,
ordered according to an increasing value of $\langle z \rangle$ (the first excited Bloch band corresponds to energies greater than the optical depth $U$ and is therefore not trapped).
\begin{table}
  \begin{center}
	  \begin{tabular}{c|c|c|c}
		  $n$ & $\langle z \rangle$ & $E_{n}-E_{n-1}$ & Perfect surface\\
		  \hline
		  $1$  & $0.799$  & $E_{1}=-0.0709$ \\
		  $2$  & $1.006$  & $+1.9690$ \\
		  $3$  & $3.372$  & $-0.5468$ \\
		  $4$  & $4.268$  & $-0.1264$ & $-0.1371$ \\
		  $5$  & $4.681$  & $-0.0934$ & $-0.0996$ \\
		  $6$  & $4.746$  & $-0.0693$ & $-0.0804$ \\
		  $7$  & $5.617$  & $-0.0579$ & $-0.0722$ \\
		  $8$  & $6.881$  & $-0.0637$ & $-0.0703$ \\
		  $9$  & $7.962$  & $-0.0679$ & $-0.0701$ \\
		  $10$ & $8.985$  & $-0.0692$ & $-0.0701$ \\
		  $11$ & $9.994$  & $-0.0696$ & $-0.0701$ \\
		  $12$ & $10.998$ & $-0.0698$ & $-0.0701$ \\
		  $13$ & $12.001$ & $-0.0700$ & $-0.0701$ \\
		  $14$ & $13.002$ & $-0.0700$ & $-0.0701$ \\
		  $15$ & $14.003$ & $-0.0700$ & $-0.0701$ \\
		  $16$ & $15.003$ & $-0.0701$ & $-0.0701$ \\
		  $17$ & $16.003$ & $-0.0701$ & $-0.0701$ \\
		  \vdots & \vdots & \vdots & \vdots
      \end{tabular}	
  \end{center}
  \caption{SMWSS for a lattice depth $U=3$ $E_r$ ordered according to their mean distance to the surface $\langle z \rangle$. Energy
  intervals are given in units of $E_{r}$. Further analysis (see text) shows that Surface-modified Wannier-Stark states begin at
  $n=3$ while the first two states are atom-surface bound states. The last column refers to the energy intervals of an infinite
  potential surface (\emph{i.e.} perfect surface).}
  \label{tab:smwss}
\end{table}
The closest SMWSS are modified very strongly by the presence of the surface, which reflects on the lack of regularity characteristic of a Wannier-Stark ladder. As $\langle z \rangle$ increases though, we progressively recover a usual
Wannier-Stark ladder spaced by the Bloch energy $h \nu_B$ and integer values of $\langle z \rangle$.

For the purpose of the FORCA-G experiment, we are mostly interested in the states closest to the surface.
\begin{figure}[h]
  \centering
    \includegraphics[width=.45\textwidth]{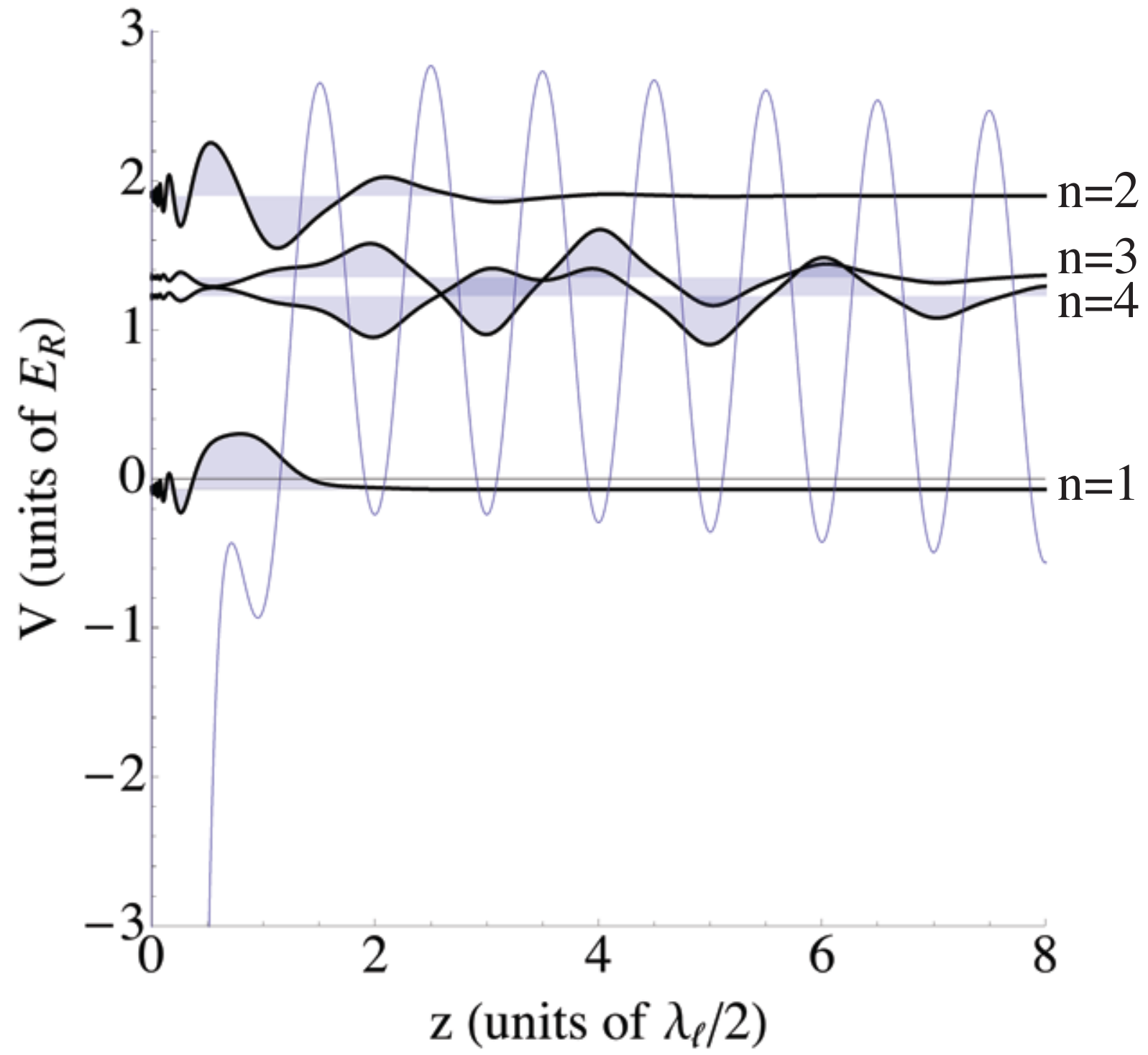}
  \caption{(color online) Wavefunctions for the first four SMWSS $\psi_{n}(z)$ according to Table~\ref{tab:smwss}. As it is customary, the vertical offset of the wavefunctions correspond to their respective energies.}
  \label{fig:Figures_WFs}
\end{figure}
We show in Figure~\ref{fig:Figures_WFs} the profile of the real wavefunctions corresponding to the first four SMWSS according to
Table~\ref{tab:smwss}. 
The probability amplitudes of the first two states exhibit very rapid oscillations within the Lennard-Jones well, while they are
vanishingly small ouside this well.
On the other hand, already
the states $n=3$ and $n=4$ are well spread along the optical potential as the ordinary Wannier-Stark states would. The tail of their
wavefunctions still show some oscillations caused by the Lennard-Jones potential.

\subsection{Dependence upon the Lennard-Jones parameters} 
\label{sub:dependence_upon_the_surface_potential}

While modeling the short-range potential, our largest uncertainty lies in the unknown shape of the potential well.
Although we have used a known analytical form which correctly converges towards the Casimir-Polder potential, the actual short-range potential may differ from the Lennard-Jones form~\cite{Sedlacek2015}.
It is therefore crucial to study the
dependence of our results upon the free parameters in $V_{\text{s}}^{\text{LJ}}(z)$.
Let us recall some general features of the bound states of a $12-3$ Lennard-Jones potential. Having a finite depth $D$
and vanishing sufficiently fast as $z\to \infty$, the potential $V_{\text{s}}^{\text{LJ}}(z)$ given by
Equation~\eqref{eq:Vsurf} possesses a \emph{finite} number of bound states. Those states represent vibrational states for
an atom bound to the surface and are therefore indexed with an integer vibrational quantum number $v$ starting with $v=0$
for the ground state. When the total number of bound states supported by a potential well is unknown (\emph{e.g.} due to
uncertainties on the dissociation energy $D$) it is customary to label the least bound states as $v=-1$, the second
least bounded states as $v=-2$ and so on. To a very good approximation, the position of the few least bound states only
depends on the asymptotic behavior of the potential as $z\to \infty$ and on a non-integer effective vibrational
quantum number at dissociation, $v_{D}$, which varies between $0$ and
$-1$~\cite{LeRoy1970}. By decreasing continuously the depth of the potential the states $v=-1$,
$v=-2$ will be eventually expelled to the continuum.
From those considerations we see that, as far as the few least bound states are concerned, the exact shape of the
potential energy well is not important. 
In our case, the effective vibrational quantum number
$v_{D}$
can be varied by simply decreasing the depth $D$ of our $12-3$ Lennard-Jones model. Owing to the
Equation~\eqref{eq:related}, the dissociation energy $D$ is decreased by increasing the equilibrium atom-surface distance
$z_{0}$ as $D(\text{eV})\approx 0.36 z_{0}^{-3}(\text{\AA})$.

\begin{figure}[htbp]
  \begin{center}
    \includegraphics[width=0.45\textwidth]{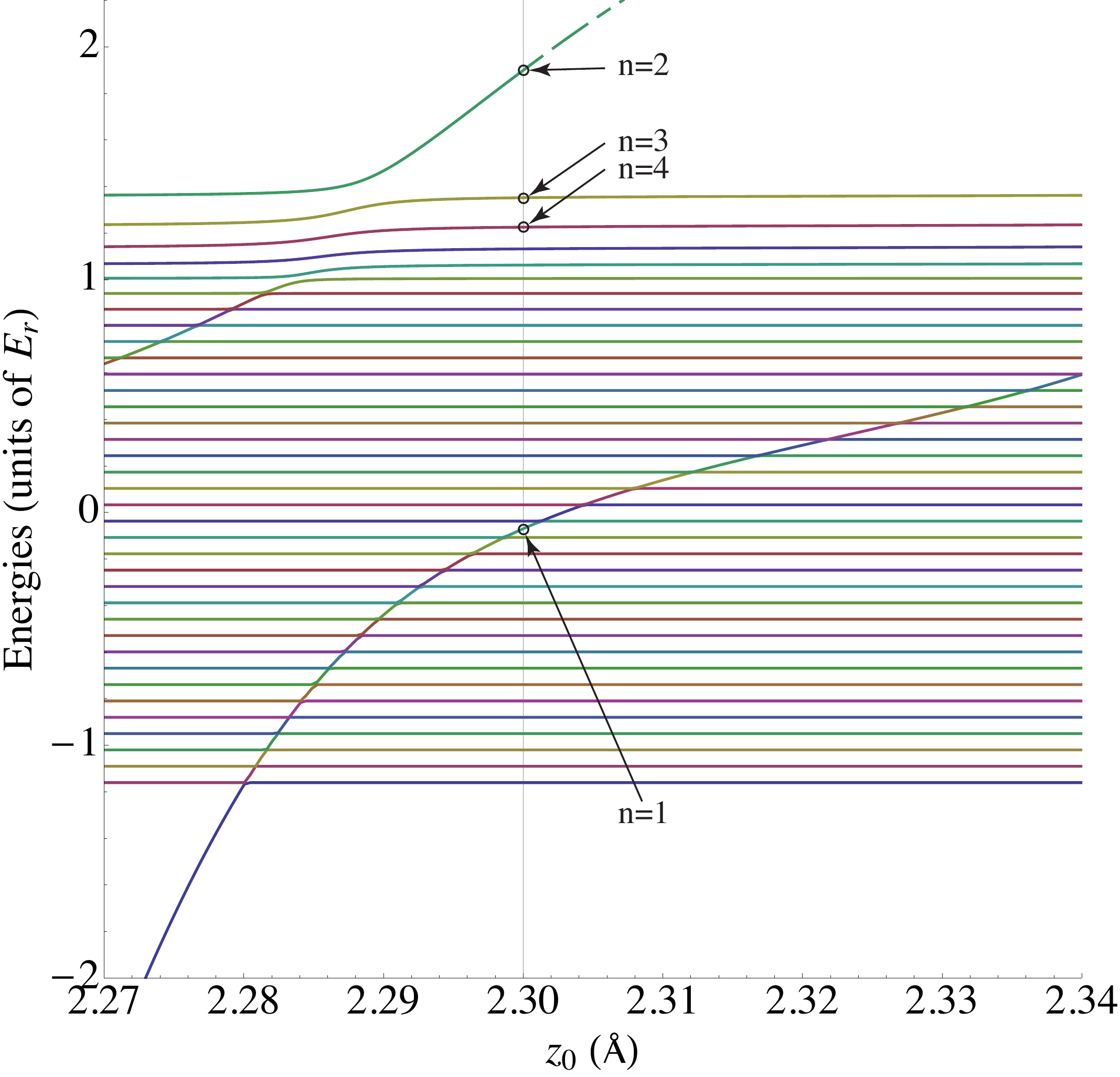}
  \end{center}
  \caption{(color online) Calculated energies of the SMWSS as a function of the distance $z_{0}$ in the Lennard-Jones potential, at constant $C_{3}$ coefficient. The first four states depicted in Figure~\ref{fig:Figures_WFs} and calculated for $z_{0}=2.3$ \AA$\,$ are indicated by arrows.}
  \label{fig:energiesZ0}
\end{figure}

We show in Figure~\ref{fig:energiesZ0} the energies of the SMWSS as a function of $z_{0}$ or, equivalently, as a function of decreasing dissociation energy $D$. In the first place, one sees that the
states $n=1$ and $n=2$ have a very different behavior compared to all the others. The position of those states
depends critically upon the dissociation energy $D$. As such, it is clear that
the two SMWSS states $n=1$ and $n=2$ are basically the last two bound states $v=-2$ and $v=-1$, respectively, of the
short-range potential. As the equilibrium distance $z_{0}$ increase, the energies of the states $n=1$ and $n=2$
increases and they cross all the other states. Nonetheless there must be avoided crossings since all those states result from
the diagonalization of the Hamiltonian operator.

On the other hand, the energies of the states starting from the $n=3$ are very much independent of the parameters used in
the short-range surface potential $V_{\text{s}}^{\text{LJ}}(z)$ except near an avoided crossing with a bound atom-surface
state. From Figure~\ref{fig:energiesZ0} we conclude that the state $n=3$ can be considered the first surface-modified
Wannier-Stark states. 
The coupling between the few first SMWSS and the atom-surface bound states quickly vanishes as $n$ increases owing to the
vastly different mean atom-surface distance $\langle z \rangle$. This leads to negligible avoided crossings between
the state $n=2$ and already the state $n=7$. Far from any avoided crossings, the SMWSS are still influenced by the surface. 
At $z_{0}=2.3$ \AA$\,$ it is shown in Table~\ref{tab:smwss} that the
energy interval between successive states is not equal to the Bloch frequency for the first Wannier-Stark states. 

It is also illustrative to compare our results with those obtained from the modeling of the short-range potential with that of a perfect surface,
\begin{equation}
  \label{eq:perfect}
  V_{\text{s}}(z)=\begin{cases}
  +\infty &z<0\\
  0 & z>0
  \end{cases}.
\end{equation}
In the first place, the repulsive part of the Lennard-Jones potential plays the role
of an infinite potential wall. 
However, in the case of an infinite potential surface the wavefunction have a different behavior at $z=0$. In particular,
the wavefunction vanishes monotically as $z \to 0$~\cite{Pelisson2013,Messina2011} whereas it oscillates very rapidly within a Lennard-Jones potential.
Obviously, a major drawback of an infinite potential surface is the total absence of bound atom-surface states. Values of the corresponding energy intervals can be found in Table~\ref{tab:smwss}.


\subsection{Simulated Raman spectrum} 
\label{sub:simulated_raman_spectrum}

The experimental setup of the FORCA-G is detailed \emph{e.g.} in reference~\cite{Beaufils2011}. In it, two counterpropagating Raman lasers operating at $\lambda=780$ nm drive coherent transitions between the ground $| 5 ^{2}S_{1/2},F=1,m_{F}=0\rangle$
and excited $| 5 ^{2}S_{1/2},F=2,m_{F}=0\rangle$ hyperfine levels of trapped $^{87}$Rb atoms. Those transitions can involve different SMWSS with a probability proportionnal to the generator of translations along the $z$-axis, $\langle \psi_{n}|e^{i k_{\text{eff}}z}| \psi_{m} \rangle$, with $k_{\text{eff}}\approx 4\pi/(780$ nm).

\begin{figure}[h]
  \centering
    \includegraphics[width=.45\textwidth]{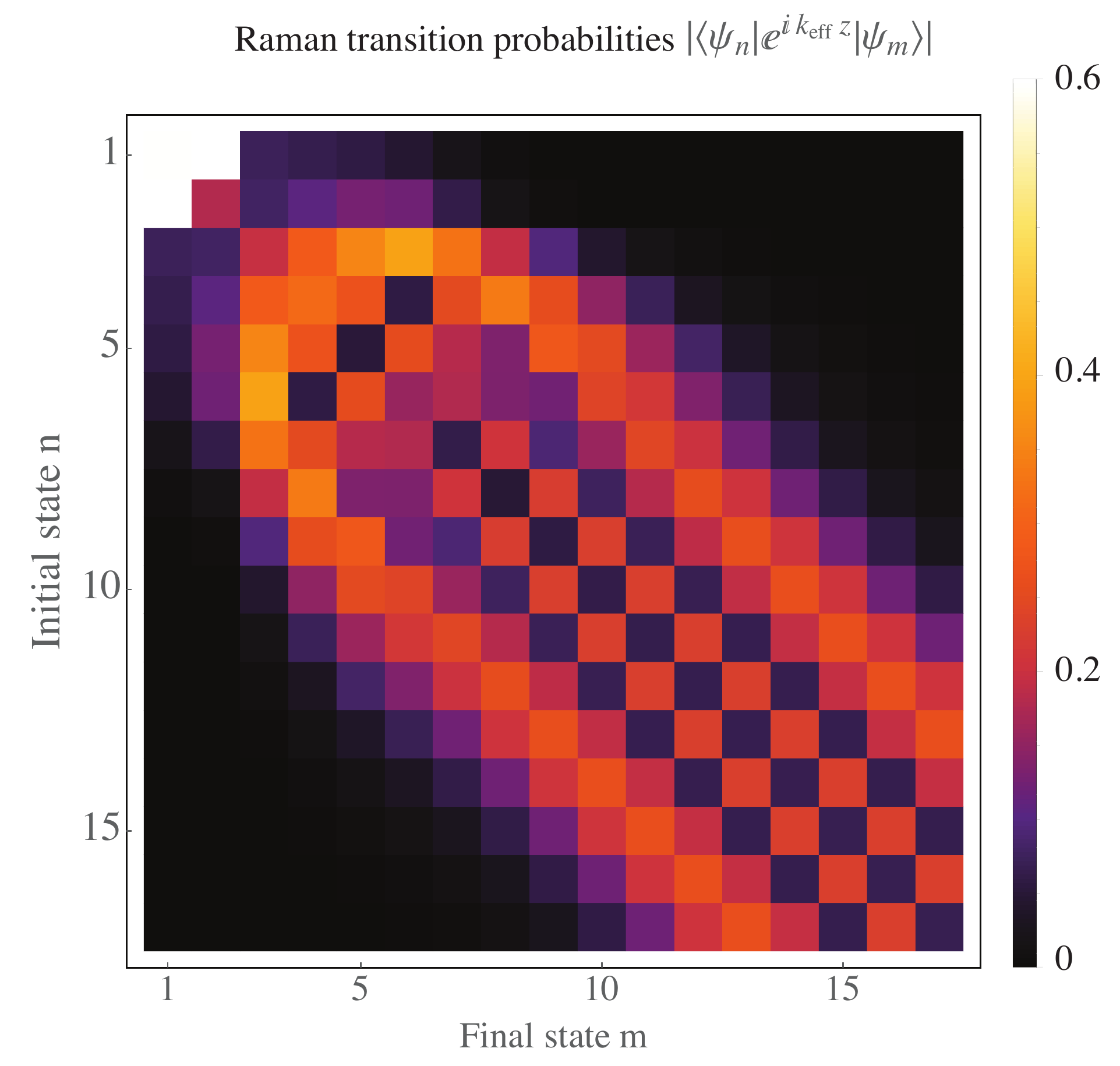}
  \caption{(Color online) Raman transition probabilities between an initial SMWSS $\psi_{n}(z)$ and a final state $\psi_{m}(z)$.}
  \label{fig:Figures_SMWSS-Proba}
\end{figure}

We show in Figure~\ref{fig:Figures_SMWSS-Proba} the Raman transition probabilities between the states presented in Table~\ref{tab:smwss}. The first two states, $\psi_1(z)$ and $\psi_2(z)$, which are the atom-surface bound states are only weakly coupled to the surface-modified Wannier-Stark states but strongly coupled to each other. We can see the smooth evolution of the SMWSS towards “regular”, unmodified Wannier-Stark states whose transition probabilities become a function of $|n-m|$ only. For a lattice depth of 3 $E_{r}$, a given state $\psi_{n}(z)$ roughly couples to states up to $n \pm 6$.

With a low-density atomic cloud like in reference~\cite{Beaufils2011}, some $10^4$ lattice sites are populated and the Raman spectrum is dominated by transitions involving unmodified Wannier-Stark states. When the frequency difference between the two Raman lasers, $\nu_\text{R}=\nu_{\text{R}_{1}}-\nu_{\text{R}_{2}}$, is scanned around the rubidium hyperfine splitting $\nu_{\text{HFS}}$, this leads to a simple spectrum with lines at integer numbers of the Bloch frequency $\nu_{B}=h^{-1}mg \lambda_l /2 \approx 568.5$ Hz. One could imagine an experiment with a much more dense atomic sample with a size of a few microns where the contribution from the SMWSS would be visible. 
\begin{figure}[h]
  \centering
    \includegraphics[width=.45\textwidth]{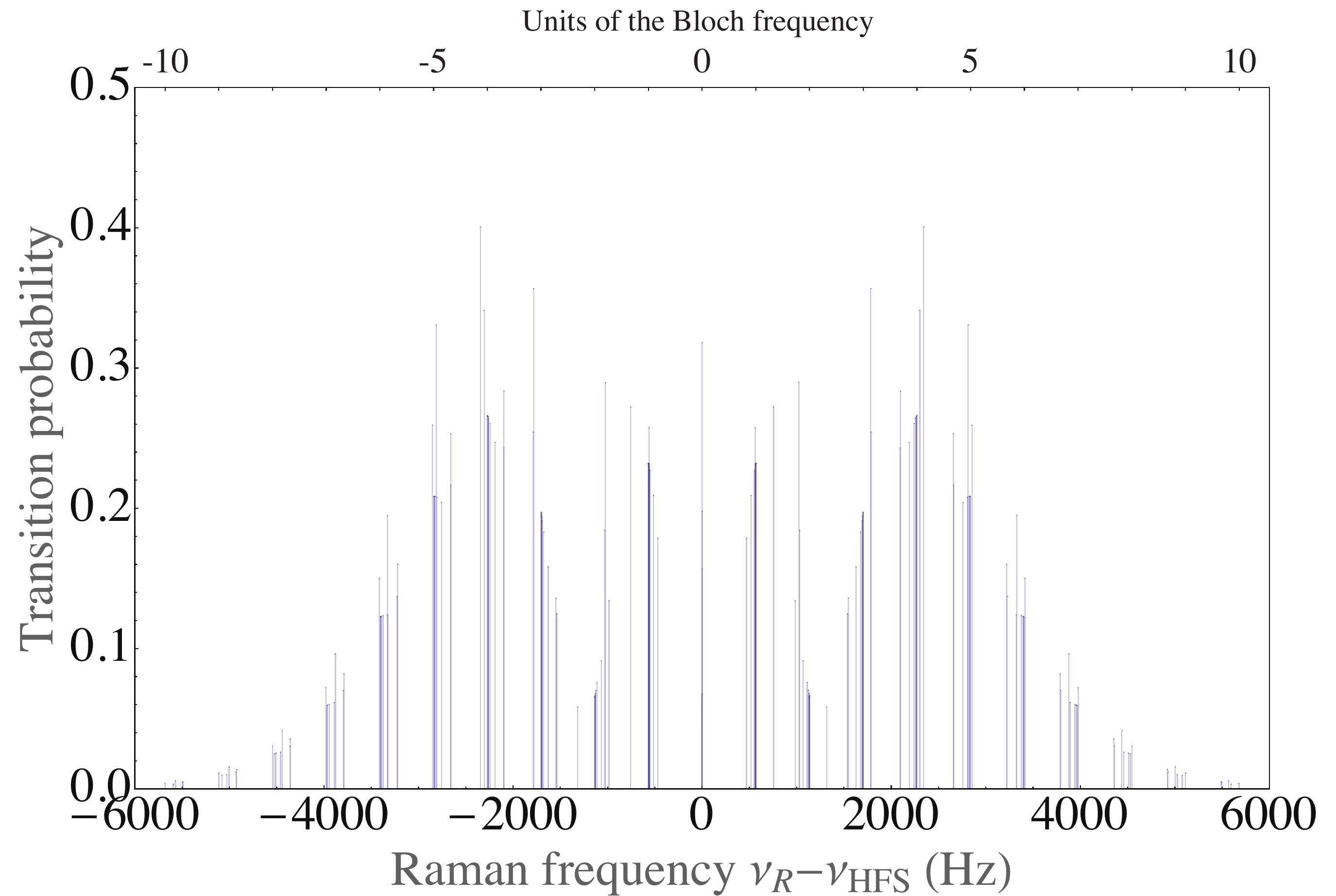}
  \caption{Raman stick-spectrum involving the states in Table~\ref{tab:smwss}. Lines involving the atom-surface bound states $\psi_1 (z)$ and $\psi_2 (z)$ are not shown.}
  \label{fig:Figures_SMWSS-Spectrum}
\end{figure}
We show in Figure~\ref{fig:Figures_SMWSS-Spectrum} the simulated Raman stick-spectrum (spectrum without line shapes) for the states
listed in Table~\ref{tab:smwss}. As we have shown in Figure~\ref{fig:energiesZ0}, the position of the atom-surface bound states
$\psi_1 (z)$ and $\psi_2 (z)$ is largely unknown. Therefore, we do not show their contributions in the spectrum of
Figure~\ref{fig:Figures_SMWSS-Spectrum}. The energies of those atom-surface bound states will appear as additional lines in the Raman
spectrum.
As expected, the departure from the regular Wannier-Stark ladder generates many lines. Those lines have the tendency to bundle up
around integer numbers of the Bloch frequency though. Recently, a relative sensibility of $4 \times 10^{-6}$ at $1$
s on the measure of the Bloch frequency has been demonstrated using a Ramsey-type interferometry~\cite{Hilico2015}. Such a
sensibility would in principle allow to resolve the lines presented in figure~\ref{fig:Figures_SMWSS-Spectrum}.



\subsection{Determination of the Casimir-Polder potential} 
\label{sub:determination_of_the_casimir_polder_potential}

Up to now, the Casimir-Polder potential has been kept constant to its calculated value in section~\ref{sec:atom_surface_potential}.
The aim of the FORCA-G experiment is to determine this Casimir-Polder potential from a recorded Raman spectrum. Thus, we have to
know how the Raman spectrum changes when one changes the Casimir-polder potential. For other references related to the use of Bloch
oscillations in order to measure the coefficients in the Casimir-Polder potential, see \emph{e.g.}~\cite{Carusotto2005,Sorrentino2009}. In the following, we
focus on the van der Waals coefficient $C_{3}$.

\begin{figure}[h]
  \centering
    \includegraphics[width=.45\textwidth]{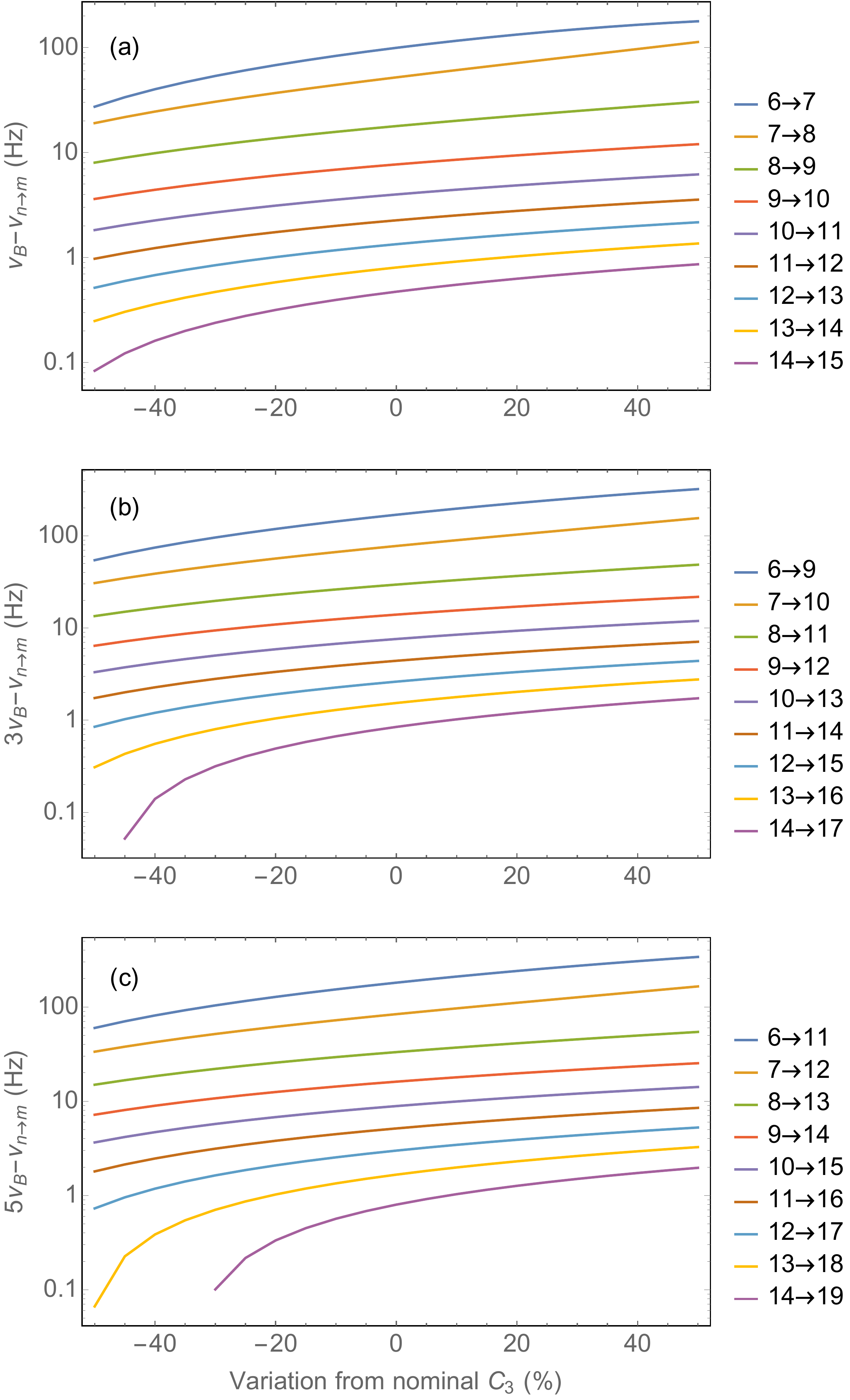}
  \caption{(Color online) Change in Raman transition frequencies $\nu_{n \to m}$, as a function of the van der Waals coefficient $C_{3}$, for selected states. (a): $|m-n|=1$ transitions. (b): $|m-n|=3$ transitions. (c): $|m-n|=5$ transitions.}
  \label{fig:spectrumVarC3}
\end{figure}

We show in figure~\ref{fig:spectrumVarC3} the change in Raman transition frequencies $\nu_{n \to m}=h^{-1}\left(E_{n}-E_{m}\right)$
when the $C_{3}$ coefficient is allowed to vary from its nominal value of $3.28$ a$_{0}^{3}$eV calculated in
section~\ref{sec:atom_surface_potential}. We present in figure~\ref{fig:spectrumVarC3} selected transitions involving
$|n-m|=1$, $3$ and $5$ and selected states $n \geq 6$. A precise analysis of the position of those lines with respect to integer
values of the Bloch frequency will allow the van der Waals coefficient $C_{3}$ to be determined. In fact, from an absolute
uncertainty of $20$ mHz~\cite{Hilico2015} on the determination of Raman transition frequencies, we infer a relative uncertainty
$\delta C_{3}/C_{3}$ on the van der Waals coefficient ranging from $10^{-2}$ to $10^{-4}$.



\section{Conclusion}

We have calculated the energies of atoms trapped in a 1D vertical optical lattice taking into account the interaction between those
atoms and the mirror used to realize the lattice. We have found that, in the range of energy of a few recoil energy $E_{r}$,
loosely bound atom-mirror states appear as additional levels among an otherwise surface-modified Wannier-Stark ladder.
The energies of those loosely bound atom-mirror states depend critically on the details of the adsorption atom-surface
potential. Atomic interferometry involving those loosely bound atom-mirror states will shed light on the adsorption dynamics of
rubidium atoms on mirrors. The close surface-modified Wannier-Stark states correspond to optically trapped atoms which
nevertheless have a significant probability of being adsorbed by the mirror.

\section{Acknowledgements} 
\label{sec:acknoledgments}
We thank Franck Pereira dos Santos and Peter Wolf for stimulating discussions. Financial support from ANR-10-IDEX-0001-02-PSL and
ANR-13-BS04–0003-02 is gratefully acknowledged.


\begin{thebibliography}{99}
%
%
%
\bibitem{Clade2006} P. Cladé, E. de Mirandes, M. Cadoret, S. Guellati-Khélifa, C. Schwob, F. Nez, L. Julien and F. Biraben, \emph{Determination of the Fine Structure Constant Based on Bloch Oscillations of Ultracold Atoms in a Vertical Optical Lattice}, Phys. Rev. Lett. \textbf{96}, 033001 (2006).
\bibitem{Takamoto2005} M. Takamoto, F.-L. Hong, R. Higashi and H. Katori \emph{An optical lattice clock}, Nature (London) \textbf{435}, 321 (2005).
\bibitem{Impens2014} F. Impens, C. Ccapa Ttira, R. O. Behunin and P. A. Maia Neto, \emph{Dynamical local and nonlocal Casimir atomic phases}, Phys. Rev. A \textbf{89}, 022516 (2014).
\bibitem{Wolf2007} P. Wolf, P. Lemonde, A. Lambrecht, S. Bize, A. Landragin, A. Clairon, \textit{From optical lattice clocks to the measurement of forces in the Casimir regime}, Phy. Rev. A \textbf{75}, 063608 (2007).
\bibitem{Casimir1948} H. B. G. Casimir and D. Polder, \emph{The Influence of Retardation on the London-van der Waals Forces}, Phys Rev \textbf{73}, 360 (1948).
\bibitem{Lifshitz1956} E. M. Lifschitz \emph{The Theory of Molecular Attractive Forces between Solids}, Soviet Physics \textbf{2}, 73 (1956).
\bibitem{Messina2011} R. Messina, S. Pelisson, M.-C. Angonin, and P. Wolf, \emph{Atomic states in optical traps near a planar surface,} Phys. Rev. A \textbf{83}, 052111 (2011).
\bibitem{Pelisson2013} S. Pelisson, R. Messina, M. C. Angonin and P. Wolf  \emph{Lifetimes of atoms trapped in an optical lattice in proximity of a surface.}, Phys. Rev. A \textbf{88}, 013411 (2013).
\bibitem{Lambrecht2006} A. Lambrecht, P. A. Maia Neto and S. Reynaud, \emph{The Casimir effect within scattering theory}, New J. Phys. \textbf{8}, 243 (2006).
\bibitem{Dufour2013} G. Dufour, A. Gérardin, R. Guérout, A. Lambrecht, V. V. Nesvizhevsky, S. Reynaud and A. Yu Voronin \emph{Quantum reflection of antihydrogen from Casimir potential above matter slabs} Phys. Rev. A \textbf{87}, 012901 (2013).
\bibitem{Derevianko2010} A. Derevianko, S. G. Porsev, J. F. Babb, \emph{Electric dipole polarizabilities at imaginary frequencies for hydrogen, the alkali-metal, alkaline-earth and noble gas atoms}, Atomic Data and Nuclear Data Tables \textbf{96} 323-331 (2010).
\bibitem{Ingold2015} G.-L. Ingold and A. Lambrecht \emph{Casimir effect from a scattering approach}, Am. J. Phys. \textbf{83}, 156 (2015).
\bibitem{Sedlacek2015} J. A. Sedlacek, E. Kim, S. T. Rittenhouse, P. F. Weck, H. R. Sadeghpour and J. P. Shaffer, \emph{Electric Field Cancellation on Quartz by Rb Adsorbate-Induced Negative Electron Affinity}, Phys. Rev. Lett. \textbf{116}, 133201 (2016).
\bibitem{Siegert1939} A. J. F. Siegert, \emph{On the derivation of the dispersion relation for nuclear reactions}, Phys. Rev. \textbf{56}, 750 (1939).
\bibitem{LeRoy1970} R. J. LeRoy and R. B. Bernstein, \emph{Dissociation Energy and Long-Range Potential of Diatomic Molecules from Vibrational Spacings of Higher Levels}, J. Chem. Phys. \textbf{52}, 3869 (1970).
\bibitem{Beaufils2011} Q. Beaufils, G. Tackmann, X. Wang, B. Pelle, S. Pelisson, P. Wolf, and F. Pereira dos Santos \emph{Laser Controlled Tunneling in a Vertical Optical Lattice}, Phys. Rev. Lett. \textbf{106}, 213002 (2011)
\bibitem{Hilico2015} A. Hilico, C. Solaro, M.-K. Zhou, M. Lopez, F. Pereira dos Santos \emph{Contrast decay in a trapped-atom interferometer}, Phys. Rev. A \textbf{91}, 053616 (2015)
\bibitem{Carusotto2005} I. Carusotto, L. Pitaevskii, S. Stringari, G. Modugno and M. Inguscio \emph{Sensitive Measurement of Forces at the Micron Scale Using Bloch Oscillations of Ultracold Atoms}, Phys. Rev. Lett. \textbf{95}, 093202 (2005)
\bibitem{Sorrentino2009} F. Sorrentino, A. Alberti, G. Ferrari, V. V. Ivanov, N. Poli, M. Schioppo and G. M. Tino \emph{Quantum sensor for atom-surface interactions below $10\mu$m} Phys. Rev. A \textbf{79}, 013409 (2009)
\end{thebibliography}
\end{document}